\documentclass[%
superscriptaddress,
preprint,
 amsmath,amssymb,
aps,
pra,
]{revtex4-2}

\usepackage[english]{babel}
\usepackage{graphicx}
\usepackage{dcolumn}
\usepackage{bm}

\usepackage{braket}
\usepackage{amsmath,amssymb}

\usepackage{hyperref}
\usepackage[capitalize]{cleveref}
\crefname{section}{Sec.}{Secs.}

\usepackage{booktabs} 
\usepackage{siunitx}  

\usepackage{xcolor}

\usepackage{acro}
\DeclareAcronym{as}{short = \textit{as},long = attosecond}
\DeclareAcronym{wp}{
  short = WP,
  long  = wave packet,
}
\DeclareAcronym{tdpt}{
  short = TDPT,
  long  = Time-dependent Perturbation Theory,
}
\DeclareAcronym{isxrs}{
  short = ISXRS,
  long  = Impulsive Stimulated X-ray Raman Scattering,
}
\DeclareAcronym{xfel}{
  short = XFEL,
  long  = X-ray Free Electron Laser,
}
\DeclareAcronym{rixs}{
  short = RIXS,
  long  = Resonant Inelastic X-ray Scattering,
}
\DeclareAcronym{eom-cc}{
  short = EOM-CC,
  long  = Equation-of-Motion Coupled-Cluster,
}
\DeclareAcronym{1rdm}{
  short = 1-RDM,
  long  = One-particle Reduced-Density-Matrix,
}

\DeclareAcronym{hf}{
  short = HF,
  long  = Hartree-Fock,
}

\begin{document}

\title{Preparation and control of electronic wave packets in neutral molecules via attosecond x-ray processes}

\author{Emanuele Rossi}
\altaffiliation[Current address: ]{DTU Chemistry, Technical University of Denmark, Kemitorvet 207, 2800 Kongens Lyngby, Denmark}
\affiliation{Department of Physics, University of Hamburg, Hamburg, Germany}
\affiliation{Center for Free-Electron Laser Science CFEL, Deutsches Elektronen-Synchrotron DESY, Hamburg, Germany}
\affiliation{The Hamburg Centre for Ultrafast Imaging, Hamburg, Germany}
\author{Stasis Chuchurka}
\altaffiliation[Current address: ]{Institute of Physical Chemistry (IPC), Friedrich Schiller University Jena, Jena, Germany}
\affiliation{Center for Free-Electron Laser Science CFEL, Deutsches Elektronen-Synchrotron DESY, Hamburg, Germany}
\affiliation{Department of Physics, University of Hamburg, Hamburg, Germany}
\author{Kaushik D. Nanda}
\altaffiliation[Current address: ]{Q-Chem, Inc. 6601 Owens Drive, Suite 240, Pleasanton, CA 94588}
\affiliation{Department of Chemistry, University of Southern California, Los Angeles, California 90089, USA}
\author{Anna I. Krylov}
\affiliation{Department of Chemistry, University of Southern California, Los Angeles, California 90089, USA}
\author{Nina Rohringer}
\email{nina.rohringer@desy.de}
\affiliation{Center for Free-Electron Laser Science CFEL, Deutsches Elektronen-Synchrotron DESY, Hamburg, Germany}
\affiliation{Department of Physics, University of Hamburg, Hamburg, Germany}
\affiliation{The Hamburg Centre for Ultrafast Imaging, Hamburg, Germany}
\author{Robin Santra}
\email{robin.santra@cfel.de}
\affiliation{Center for Free-Electron Laser Science CFEL, Deutsches Elektronen-Synchrotron DESY, Hamburg, Germany}
\affiliation{Department of Physics, University of Hamburg, Hamburg, Germany}
\affiliation{The Hamburg Centre for Ultrafast Imaging, Hamburg, Germany}

\date{\today\clearpage}

\begin{abstract}
We present a perturbative framework for computing the dynamics of an electronic wave packet launched by an attosecond x-ray pulse in a neutral molecule. The x-ray pulse excites the molecule via both x-ray absorption and \ac{isxrs}, coherently populating both the core-excited states and valence-excited states. We describe the electronic structure within the Equation-of-Motion Coupled-Cluster framework, adopting a compact representation of the sum over states expression characterizing the second-order perturbative description of \ac{isxrs}. We study the coherent electronic dynamics in OCS and oxazole utilizing the time-dependent difference electron density, decomposing it in terms of its perturbative components. While the core-excited components dominate the initial stages of the dynamics, the valence-excited components dominate the electronic dynamics on a longer time scale. The pulse polarization controls the symmetry of the states included in the wave packet. We show how the spatial symmetry of the states plays a role in shaping the spatial properties of charge migration. The atom-specificity of the \ac{isxrs} process translates directly into the initial localization of the wave packet. This determines the starting point of charge migration, shaping its subsequent evolution.
\end{abstract}

\maketitle

\section{Introduction}
The emergence of light sources capable of producing ultrashort pulses has allowed to study the dynamics of a photochemical process with ever higher time resolution. Pulses of femtosecond duration gave access to the time scale of nuclear motion; this enabled to reconstruct and control the dynamics of a photochemical reaction on a femtosecond time scale, establishing the field of femtochemistry \cite{Femto_chem-Zewail}. 

The introduction of attosecond pulses \cite{Hu_2026,atto_pulses_1} pushed the time resolution to that of electronic motion, introducing an 'electronic time scale' to the study of a photochemical reaction \cite{Remacle_Levine_2006,Lepine_Ivanov_Vrakking_2014}. The intramolecular distribution of the electrons plays a crucial role in shaping the molecular bonding network: owing to the coupling between the nuclei and the electrons, a reorganization of the electronic distribution can influence the nuclear dynamics and potentially route the chemical process on a particular pathway \cite{Weitzel_2007,charge_dir_react_1,charge_dir_react_2, Kuleff_Cederbaum_2014}. Based on this hypothesis, the idea of 'attochemistry' has emerged \cite{Merritt_Jacquemin_Vacher_2021}, which proposes to control a chemical process using the properties of an attosecond pulse to steer the electronic dynamics. 

\begin{figure}
    \centering
    \includegraphics[width=\textwidth]{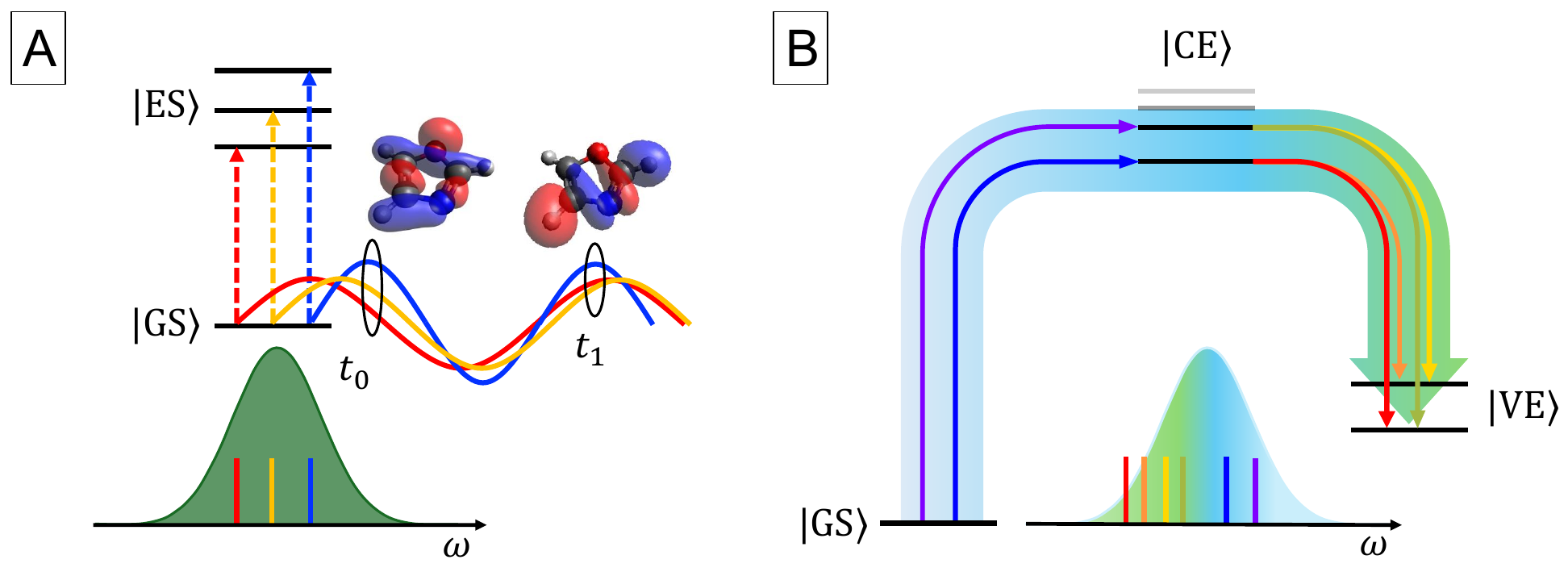}
    \caption{Preparation of an electronic \ac{wp}, charge migration, and schematic description of the \ac{isxrs} process. (A) Coherent excitation of the system from the ground state $\ket{\text{GS}}$ to a manifold of excited states $\ket{\text{ES}}$. The time evolution of the coherent superposition corresponds to the motion of the electronic density known as charge migration. (B) Schematic view of the \ac{isxrs} excitation with a broadband x-ray pulse. The manifold of core-excited and valence-excited states are indicated as $\ket{\text{CE}}$ and $\ket{\text{VE}}$, respectively. The arrows connecting $\ket{\text{GS}}$ and $\ket{\text{CE}}$ refer to the x-ray absorption step; the arrows connecting $\ket{\text{CE}}$ and $\ket{\text{VE}}$ refer to the x-ray stimulated emission step. Their coloring scheme corresponds to the respective 'absorbed' and 'emitted' frequencies, indicated by the colored bars contained within the pulse bandwidth.}
    \label{fig:intro_WP_SRIXS}
\end{figure}

The reorganization of the electronic distribution underlying attochemistry is based on a purely electronic process, commonly known as charge migration \cite{Cederbaum_Zobeley_1999,Worner__2017}. As shown in Fig.~\ref{fig:intro_WP_SRIXS}A, charge migration can be launched via a broadband excitation, which prepares a coherent superposition of electronic states (i.e., an electronic \ac{wp}) \cite{Calegari_2014}. The motion of the electronic density underlying charge migration corresponds to the time evolution of the superposition. Due to the high photon energies of the attosecond pulses---which often exceed the molecular ionization threshold---the pioneering studies in attochemistry considered a photoionization process to prepare the electronic \ac{wp} \cite{Cederbaum_Zobeley_1999,Kuleff_Cederbaum_2014,Calegari_2014,Kraus_2015,Folorunso_2023,Fransen_Gomez_Vacher_2025,Golubev_Kuleff_2015}. However, since fundamental photochemical reactions (such as those involved in vision or photosynthesis \cite{Glusac_2016}) rely on photoexcitation, the focus of recent studies has moved towards electronic dynamics in neutral molecules \cite{Silane_neutral_2022,Open_attoquestions_2023}.

Contributing to this direction of research, we theoretically study charge migration in a neutral molecule, preparing the electronic \ac{wp} with a broadband, attosecond x-ray pulse. In particular, we consider an attosecond x-ray pulse delivered by an \ac{xfel} \cite{LCLS-AS_pulses}. The nature of this pulse makes it possible to coherently populate the electronic states of the molecule in an atom-specific way \cite{Stohr1992}. Here, we populate both the core-excited and the valence-excited states. The latter can be populated by exploiting the high intensity of an \ac{xfel} pulse, which gives access to nonlinear spectroscopic techniques such as \ac{isxrs} \cite{Tanaka_Mukamel_2002,Harbola_Mukamel_2009,Biggs_Mukamel_2012,Rohringer_rev_2019}. This technique was experimentally demonstrated both in atomic \cite{Weninger_2013} and molecular systems \cite{Kimberg_2016,ONeal_2020,Oliver_2024,Li_2025}. A schematic description of this process (see \cref{fig:intro_WP_SRIXS}B) relies on a sequence of two steps: an x-ray absorption (light-blue background), which excites the molecule from its ground state to a manifold of core-excited states; and an x-ray stimulated emission (green background), which radiatively de-excites the molecule to a manifold of valence-excited states. Performing an \ac{isxrs} excitation using an x-ray pulse with a sufficiently large bandwidth, so that both the absorbed and emitted photon energies are contained within its spectrum, prepares a coherent superposition of the ground state and valence-excited states. The properties of such a superposition are atom-specific and can be regulated by tuning the x-ray pulse \cite{Schweigert_Mukamel_2007,Healion_Mukamel_2012,Yong_Mukamel_2021,Fouda_2021}.

The accurate modeling of the excitation and the electronic dynamics represents a theoretical challenge for electronic-structure theory and a fundamental component in the development of attochemistry. A wealth of time-dependent electronic-structure methods have been developed to tackle this challenge. Some examples are: time-dependent density functional theory \cite{tddft_Runge_Gross,Véniard_Taïeb_Maquet_2001,Bruner_Lopata_2017}; time-dependent configuration-interaction methods \cite{Greenman_Santra_2010,Carlstrom_Spanner_Patchkovskii_2022,Klamroth_2006,Ulusoy_Stewart_Wilson_2018, Krause_Klamroth_Saalfrank_2007,Krause_Klamroth_Saalfrank_2007,Krause_Klamroth_Saalfrank_2005}; multi-configuration time-dependent Hartree methods \cite{Meyer_Manthe_Cederbaum_1990,Lode_Alon_2020,mctdh_Scrinzi_2005,Hochstuhl_Hinz_Bonitz_2014}; time-dependent Hartree-Fock methods \cite{Kulander_1987}; complete \cite{Sato_Ishikawa_2013} and restricted \cite{Miyagi_Madsen_2017} variants of time-dependent active-space self-consistent-field methods; time-dependent multiconfiguration self-consistent-field methods \cite{mctdh_Kato_Kono_2004}; methods based on the algebraic adiabatic construction \cite{Dreuw_Wormit_2015,ADC_Kuleff_Cederbaum_2005,ADC_Ruberti_2018}; time-dependent coupled-cluster methods \cite{Koch_Jorgensen_1990,Skeidsvoll_Balbi_Koch_2020,Kvaal_2012,Sato_TDOCC_1_2018,Sato_TDOCC_2_2021}, and time-dependent \ac{eom-cc} methods \cite{Sonk_Caricato_Schlegel_2011,Luppi_Head-Gordon_2012,Skeidsvoll_tdeomcc_2022,Balbi_Skeidsvoll_Koch_2023}. 

Here, we adopt an approach based on \ac{eom-cc} theory \cite{Bartlett_EOM_CC,Krylov_EOMCC_2008}, which provides an accurate description of the molecular electronic structure and the x-ray spectroscopic properties \cite{XABOOM}. In particular, we expand the electronic \ac{wp} in a basis of \ac{eom-cc} states, approximating the corresponding time-dependent probability amplitudes through \ac{tdpt}. The corresponding perturbative expressions contain the \ac{rixs} transition moment, which is characterized by a sum over states, formally extending over the whole molecular spectrum. 

The common approach to the calculation of the \ac{rixs} moment resorts to its truncation to a manifold of presumably dominant terms \cite{RIXS_truncated_1,RIXS_truncated_2}; while this presents practical advantages, the loss of accuracy connected to the truncation of the sum over states is difficult to assess \textit{a priori}. Here, we make use of the \ac{eom-cc} implementation by Nanda \textit{et al.} \cite{RIXS_Qchem}, which reformulates the \ac{rixs} transition moment in a closed form within damped response theory. This approach formally avoids the truncation of the sum over core-excited states. By integrating the nontruncated formulation of the \ac{rixs} transition moments in the calculation of the \ac{isxrs} probability amplitudes, we aim at improving the description of the \ac{isxrs} excitation (in the perturbative regime) compared to previous works \cite{Balbi_Skeidsvoll_Koch_2023,Yong_Mukamel_2021,Schweigert_Mukamel_2007,Healion_Mukamel_2012}, which include only a limited set of core-excited states in the manifold of intermediate states. 

Within this framework, we study charge migration launched by an attosecond x-ray pulse in a neutral molecule by observing the changes in its time-dependent difference electron density (obtained by subtracting the ground-state electron density from the time-dependent electron density). In particular, we examine the relative contributions provided by the core and valence excitations by decomposing the time-dependent difference electron density in terms of its perturbative components. We perform this analysis in carbonyl sulfide (OCS) and oxazole, studying the properties of charge migration and how they are shaped by the pulse polarization and the inner-shell edge chosen for the x-ray excitation.

The paper is organized as follows: in Sec.~\ref{sec:theoretical methods} we present the perturbative description of the electronic \ac{wp} and the details of its implementation within \ac{eom-cc} theory. In Sec.~\ref{sec:diff_dens_dec}, we present the perturbative decomposition of the time-dependent difference electron density in OCS, describing their hierarchy and spatial contributions. In Sec.~\ref{sec:OCS_migration}, we describe the valence-electron dynamics excited by \ac{isxrs} in OCS and how they are affected by the x-ray polarization. In Sec.~\ref{sec:oxazole_migration}, we describe the valence electron dynamics excited by \ac{isxrs} in oxazole and their atom-specific properties. Finally, in Sec.~\ref{sec:conclusions} we discuss our conclusions.

\section{Theoretical methods}\label{sec:theoretical methods}
In this section, we employ atomic units. Throughout, we assume the nuclei to be clamped to their equilibrium positions. We describe the dynamics of the electronic system according to the semiclassical Hamiltonian
\begin{equation}\label{eqn:Hamiltonian}
    \hat{H}(t) = \hat{H}_{\text{el}}-\hat{\bm{\mu}}\cdot\bm{E}(t).
\end{equation}
The term $\hat{H}_{\text{el}}$ is the nonrelativistic electronic Hamiltonian  \cite{ModernQC}, while $\hat{\bm{\mu}}$ is the electric-dipole operator. The classical electric field 
\begin{equation}\label{eqn:pulse}
    \bm{E}(t) = \bm{\epsilon}E(t)\cos{(\omega_f t)}
\end{equation}
represents the incoming pulse of Gaussian envelope $E(t)$, carrier frequency $\omega_f$, and polarization $\bm{\epsilon}$. Since we consider only linearly polarized x-ray pulses, $\bm{\epsilon}$ is a real-valued vector. 

We represent the \ac{wp}, $\ket{\Psi(t)}$, as a superposition of eigenstates of $\hat{H}_{\text{el}}$. In particular, we focus on the states of the neutral molecule, considering the ground state $\ket{\psi_g}$, the core-excited states $\ket{\psi_c}$, and the valence-excited states $\ket{\psi_v}$. Assuming that the resonant enhancement makes the excitation pathways dominant over the photoionization pathways, we exclude the core-ionized and valence-ionized states from the expansion. Based on these assumptions, we expand $\ket{\Psi(t)}$ as
\begin{align}\label{eqn:spectral_expansion}
    \ket{\Psi(t)}&=[1+a^{(2)}_g(t)]\ket{\psi_g}+\sum_c a_c^{(1)}(t)e^{-i(\omega_{cg}-i\frac{\Gamma_c}{2})t}\ket{\psi_c}\nonumber\\
    &+\sum_{v}a_v^{(2)}(t)e^{-i(\omega_{vg}-i\frac{\Gamma_v}{2})t}\ket{\psi_v}.
\end{align}
The superscripts added to the probability amplitudes in this expression indicate the leading order in the perturbation $\hat{\bm{\mu}}\cdot\bm{E}(t)$ to which each amplitude must be computed in order to obtain a (potentially) nonvanishing contribution. Since we consider resonant x-ray excitations, second-order corrections to the amplitudes associated with core-excited states may be neglected.

The probability amplitudes $a_c^{(1)}(t)$ and $a_v^{(2)}(t)$ are modulated by exponential factors which control their oscillation (at frequencies proportional, respectively, to the excitation energies $\omega_{cg}$ and $\omega_{vg}$) and exponential decay, which accounts for the spontaneous relaxation of the excited states. The parameter $\Gamma$ corresponds to the rate of decay, which is related to the state's mean lifetime $\tau=1/\Gamma$. We consider the same value of $\Gamma_c$ for all core-excited states, choosing a value of 0.005 a.u. (0.140 eV) for excitations at the O $K$ edge \cite{Gamma_OK}, 0.0049 a.u. (0.132 eV) for excitations at the N $K$ edge \cite{Gamma_NK}, and 0.0039 a.u. (0.105 eV) for excitations at the S $L_1$ edge \cite{Gamma_SL}. Since $\tau_v$ is much larger than $\tau_c$ \cite{Gamma_valence}, and we study the dynamics in a time window of few femtoseconds after the excitation, we consider $\Gamma_v$ equal to zero for all the valence-excited states. 

The first-order amplitudes,
\begin{equation}\label{eqn:core_exc}
    a_c^{(1)}(t) = \frac{1}{\sqrt{2\pi}}\int\limits_{0}^{\infty}d\omega\frac{e^{i(\omega_{cg}-\omega-i\frac{\Gamma_c}{2})t}}{\omega_{cg}-\omega-i\frac{\Gamma_c}{2}}E(\omega)\bm{\epsilon}\cdot\bm{\mu}_{cg},
\end{equation}
describe the coherent population of the core-excited states upon absorption of an x-ray photon. Here, $\bm{\mu}_{cg}$ represents the transition dipole moment between the ground state and the core-excited state; $E(\omega)$ corresponds to the Fourier transform of the pulse in \cref{eqn:pulse}, multiplied by the polarization vector $\bm{\epsilon}$. The second-order contributions, $a_v^{(2)}(t)$ and $a_g^{(2)}(t)$, describe the coherent population of the valence-excited states (via \ac{isxrs}) and the depopulation of the ground state upon excitation, respectively. They are given by
\begin{widetext}
\begin{equation}\label{eqn:val_exc}
    a_k^{(2)}(t) = \frac{1}{2\pi}\int\limits_{0}^{\infty}d\omega_p \int\limits_{0}^{\infty}d\omega_d\frac{e^{i(\omega_{kg}-\omega_p+\omega_d-i\frac{\Gamma_{k}}{2})t}}{\omega_{kg}-\omega_p+\omega_d-i\frac{\Gamma_{k}}{2}}E(\omega_p)E(\omega_d)M_{kg}(\omega_p,\omega_d),
\end{equation}
\end{widetext}
with $k\in{v, g}$. Here, the term $M_{kg}(\omega_p,\omega_d)$ corresponds to the second-order transition moment of the form
\begin{widetext}
\begin{equation}\label{eqn:rixs_tm}
    M_{kg}(\omega_p,\omega_d) = 
    \sum_m\Big[\frac{(\bm{\epsilon}_d\cdot\bm{\mu}_{km})(\bm{\epsilon}_p\cdot\bm{\mu}_{mg})}{\omega_{mg}-\omega_p-i\frac{\Gamma_{m}}{2}}
    +\frac{(\bm{\epsilon}_p\cdot\bm{\mu}_{km})(\bm{\epsilon}_d\cdot\bm{\mu}_{mg})}{\omega_{mg}+\omega_d-i\frac{\Gamma_{m}}{2}}\Big],
\end{equation}
\end{widetext}
where the sum over states formally extends over the whole spectrum of $\hat{H}_{\text{el}}$. This expression is commonly found in the Kramers-Heisenberg dispersion formula \cite{KH_formula} and the description of the \ac{rixs} process \cite{RIXS_review}. Owing to the resonant enhancement at x-ray frequencies, the contributions from the core-excited states play a dominant role in the sum over states in \cref{eqn:rixs_tm}. Accordingly, we limit the sum over states to a sum over core-excited states. Instead of truncating the sum over core-excited states to a manifold of core-excited states \cite{Balbi_Skeidsvoll_Koch_2023,Yong_Mukamel_2021,Schweigert_Mukamel_2007,Healion_Mukamel_2012, RIXS_truncated_1,RIXS_truncated_2}, we use its closed-form reformulation within damped response theory, as presented in Ref.~\cite{RIXS_Qchem} and implemented within Q-Chem 6.1 \cite{Q-chem_package}.

\subsection{Perturbative decomposition of the electron density}\label{sec:pert_dec_density}
To obtain a clearer description of its coherent properties, we represent the superposition state in terms of its density matrix, $\hat{\rho}(t)=\ket{\Psi(t)}\bra{\Psi(t)}$. By substituting the expansion in \cref{eqn:spectral_expansion} in the definition of $\hat{\rho}(t)$, we write the perturbative expansion of the density matrix elements as
\begin{equation}\label{eqn:dm_coeff_generic}
    \rho_{nm}^{(N+M)}(t) = a_n^{(N)}(t)a^{(M)^*}_m(t)e^{-i(\omega_{mn}-i\frac{\Gamma_m+\Gamma_n}{2})t}.
\end{equation}
Here, $N$ and $M$ represent the perturbative orders of the probability amplitudes in \cref{eqn:spectral_expansion}; their sum gives the order of the corresponding density matrix element. Having expanded the probability amplitudes to leading second order, the perturbative expansion of the density matrix elements contains terms up to fourth order. The latter expansion is however complete only up to leading second order, which we adopt as our truncation threshold. As a consequence, the populations of the valence-excited states are neglected, while those of the ground and core-excited states are considered. Within this framework, the ground-state population is given by
\begin{equation}\label{eqn:pert_dm_gg}
    \rho_{gg}(t)\approx 1+\rho_{gg}^{(2)}(t) = 1+2\Re{[a_g^{(2)}(t)]},
\end{equation}
where the second term represents the depopulation of the ground state, while the populations of the core-excited states are obtained via
\begin{equation}\label{eqn:pert_dm_cc}
    \rho_{cc}(t)\approx\rho_{cc}^{(2)}(t) = a_c^{(1)}(t)a_c^{(1)^*}(t)e^{-\Gamma_c t}.
\end{equation}
Through second order for the density matrix, the only nonvanishing coherences are those between the ground state and the core-excited states,
\begin{equation}\label{eqn:pert_dm_cg}
    \rho_{cg}(t)\approx\rho_{cg}^{(1)}(t) = a_c^{(1)}(t)e^{-i\omega_{cg}t}e^{-\frac{\Gamma_c}{2}t},
\end{equation} 
between core-excited states, 
\begin{equation}\label{eqn:pert_dm_cc'}
    \rho_{cc'}(t)\approx\rho_{cc'}^{(2)}(t) = a_c^{(1)}(t)a_{c'}^{(1)^*}(t)e^{-i\omega_{cc'}t}e^{-\Gamma_c t},
\end{equation}
and between the ground state and the valence-excited states,
\begin{equation}\label{eqn:pert_dm_vg}
    \rho_{vg}(t)\approx\rho_{vg}^{(2)}(t) = a_v^{(2)}(t)e^{-i\omega_{vg}t}e^{-\frac{\Gamma_v}{2}t}.
\end{equation}
We choose the time-dependent electron density as the observable to obtain a direct, spatially-resolved picture of the changes in the electronic distribution upon excitation of the electronic \ac{wp}. The density-based analysis can also offer a connection to the properties of spectroscopic observables \cite{Krylov_2020}. Due to the perturbative nature of the excitation, the time-dependent electron density, $\rho(\bm{r},t)$, is dominated by the (time-independent) ground-state electron density, $\rho_{gg}(\bm{r})$. To analyze the changes in the electron density due to the excitation, we consider the difference density
\begin{equation}\label{eqn:diff_dens}
    \rho_{\Delta}(\bm{r},t) = \rho(\bm{r},t) - \rho_{gg}(\bm{r}).
\end{equation}
 The perturbative expansion of the density matrix, with components as given in \cref{eqn:pert_dm_gg,eqn:pert_dm_cc,eqn:pert_dm_cg,eqn:pert_dm_cc',eqn:pert_dm_vg}, leads to the following expansion of the difference density:
\begin{widetext}
\begin{equation}\label{eqn:density_comp}
    \rho_{\Delta}(\bm{r},t) \approx \underbrace{\sum_c\rho^{(2)}_{cc}(t)\rho_{cc}(\bm{r})+\rho_{gg}^{(2)}(t)\rho_{gg}(\bm{r})}_{\Delta_{cg}(\bm{r},t)} + \underbrace{\sum_c\rho^{(1)}_{cg}(t)\rho_{cg}(\bm{r})}_{\rho_{cg}(\bm{r},t)} + \underbrace{\sum_v\rho^{(2)}_{vg}(t)\rho_{vg}(\bm{r})}_{\rho_{vg}(\bm{r},t)} + \underbrace{\sum_{c'>c}\rho^{(2)}_{cc'}(t)\rho_{cc'}(\bm{r})}_{\rho_{cc'}(\bm{r},t)}.
\end{equation}
\end{widetext}
Here, each contribution to $\rho(\bm{r},t)$ corresponds to either a population or a coherence of the density matrix. Each density matrix component acts as a time-dependent weight for the corresponding state or transition density, which is defined as a real object of the form
\begin{equation}\label{eqn:density_def}
    \rho_{nm}(\bm{r}) = \sum_{pq}\gamma_{pq}^{nm}\phi_p(\bm{r})\phi_q(\bm{r}).
\end{equation}
In this expression, the $\gamma_{pq}^{nm} = \bra{\psi_m}\hat{c}_q^{\dagger}\hat{c}_p\ket{\psi_n}$ are the state (for $n=m$) or the transition (for $n\neq m$) \ac{1rdm} elements, while $\phi_{p}(\bm{r})$ and $\phi_{q}(\bm{r})$ correspond to (real-valued) canonical molecular orbitals.

\subsection{Implementation within \ac{eom-cc} theory}
We calculate the (state and transition) electron densities in \cref{eqn:density_comp} and the transition moments in \cref{eqn:core_exc,eqn:val_exc} within \ac{eom-cc} theory, which provides an accurate framework for the calculation of x-ray transition properties \cite{XABOOM}. The calculation of the components of $\rho_{\Delta}(\bm{r},t)$ associated with the coherences requires special attention, which is related to the non-Hermitian nature of \ac{eom-cc} theory. In standard quantum mechanics, the matrix element $\mathcal{O}_{nm}$ of a generic operator $\hat{\mathcal{O}}$, is defined in terms of an orthonormal bra and ket couple, i.e., $\mathcal{O}_{nm} = \bra{\psi_n}\hat{\mathcal{O}}\ket{\psi_{m}}$. Within the non-Hermitian \ac{eom-cc} framework, biorthonormal left bras and right kets are computed, which leads to the appearance of the two distinct matrix elements \cite{Bartlett_EOM_CC}:
\begin{subequations}\label{eqn:EOM_CC-tr_prop}
    \begin{equation}\label{subeq:EOM_CC-tr_prop-a}
        \mathcal{O}_{n\leftarrow m} = \bra{\psi_n^L}\hat{\mathcal{O}}\ket{\psi_{m}^R}
    \end{equation}
    \text{and}
    \begin{equation}\label{eqn:EOM_CC-tr_prop-b}
        \Tilde{\mathcal{O}}_{m\leftarrow n} = \bra{\psi_{m}^L}\hat{\mathcal{O}}^{\dagger}\ket{\psi_n^R},
    \end{equation}
\end{subequations}
with $\mathcal{O}_{n\leftarrow m}\neq\Tilde{\mathcal{O}}^*_{m\leftarrow n}$ when $m\neq n$. For the evaluation of \cref{eqn:density_comp}, we require the Hermitian matrix element $\mathcal{O}_{nm}$, which we express in the generic form
\begin{equation}\label{eqn:EOM_CC-hermitian_matel_def}
    \mathcal{O}_{nm} = \left|{\mathcal{O}_{nm}}\right|e^{i\phi_{nm}}.
\end{equation}
To evaluate the modulus of $\mathcal{O}_{nm}$ we adopt the geometric averaging approach proposed by \citeauthor{Bartlett_EOM_CC} \cite{Bartlett_EOM_CC}, according to which
\begin{equation}\label{eqn:EOM_CC-tr_prop_sq}
    \left|{\mathcal{O}_{nm}}\right| = \sqrt{\mathcal{O}_{n\leftarrow m}\Tilde{\mathcal{O}}_{m\leftarrow n}}.
\end{equation}
We obtain the phase factor appearing in \cref{eqn:EOM_CC-hermitian_matel_def} directly from the phase of the \ac{eom-cc} transition matrix element:
\begin{equation}\label{eqn:EOM_CC-hermitian_phase}
    e^{i\phi_{nm}} = e^{i\phi_{n\leftarrow m}} = \frac{\mathcal{O}_{n\leftarrow m}}{|\mathcal{O}_{n\leftarrow m}|}.
\end{equation}
We apply this procedure to all the \ac{eom-cc} transition properties---i.e., the transition dipole moments, the \ac{1rdm}s, and second-order transition moments.

\subsection{Computational details}\label{sec:comp_details}
All the quantum-chemical calculations are performed using a developer version of the Q-Chem 6.1 software package \cite{Q-chem_package}, modified to output both $\gamma_{pq}^{m\leftarrow n}$ and $\gamma_{pq}^{n\leftarrow m}$. The ground-state equilibrium geometries of OCS and oxazole are optimized at the CCSD/aug-cc-pVDZ level of theory (see \cref{App:equilibrium_geom} for details). The calculations of the respective electronic structures are performed at the fc-CVS-EOM-EE-CCSD/6-311+G* level of theory (see \cref{App:calc_protocol} for details). For OCS, we include a total of 23 electronic states (ground state, 15 core-excited states, and 7 valence-excited states) in the spectral expansion in \cref{eqn:spectral_expansion}; for oxazole, we include a total of 51 electron states (ground state, 30 core-excited states, and 20 valence-excited states). See \cref{App:OCS_spectrum} for a detailed description of the energy spectra of OCS and oxazole. We consider a pulse bandwidth of $\Delta\omega = 8$~eV (corresponding to a pulse duration $\Delta t\approx 230$ as) sufficient to coherently populate a large portion of the excited states included in the spectral expansion. Pulses of similar duration/bandwidth in the water window are available at \ac{xfel} facilities \cite{LCLS-AS_pulses}. We consider a pulse intensity of $10^{16}~\text{W}~\text{cm}^{-2}$, which is easily achievable at \ac{xfel} facilities and is consistent with the perturbative regime required by the theory. We consider a pulse resonant with the $1s\to\pi^*$ line at the O $K$ edge and S $L_1$ edge for OCS and at the O $K$ edge and N $K$ edge for oxazole.

\section{Results}
\subsection{The difference density and its decomposition}\label{sec:diff_dens_dec}
As a first step, we focus on the analysis of $\rho_{\Delta}(\bm{r},t)$ and its perturbative decomposition in OCS. In its equilibrium configuration, OCS is linear ($C_{\infty v}$). Since it contains atoms of three different elements, OCS provides energetically well-separated inner-shell edges one can choose for the x-ray excitation. 

\begin{figure*}[t!]
\centering
    \includegraphics[width=\textwidth]{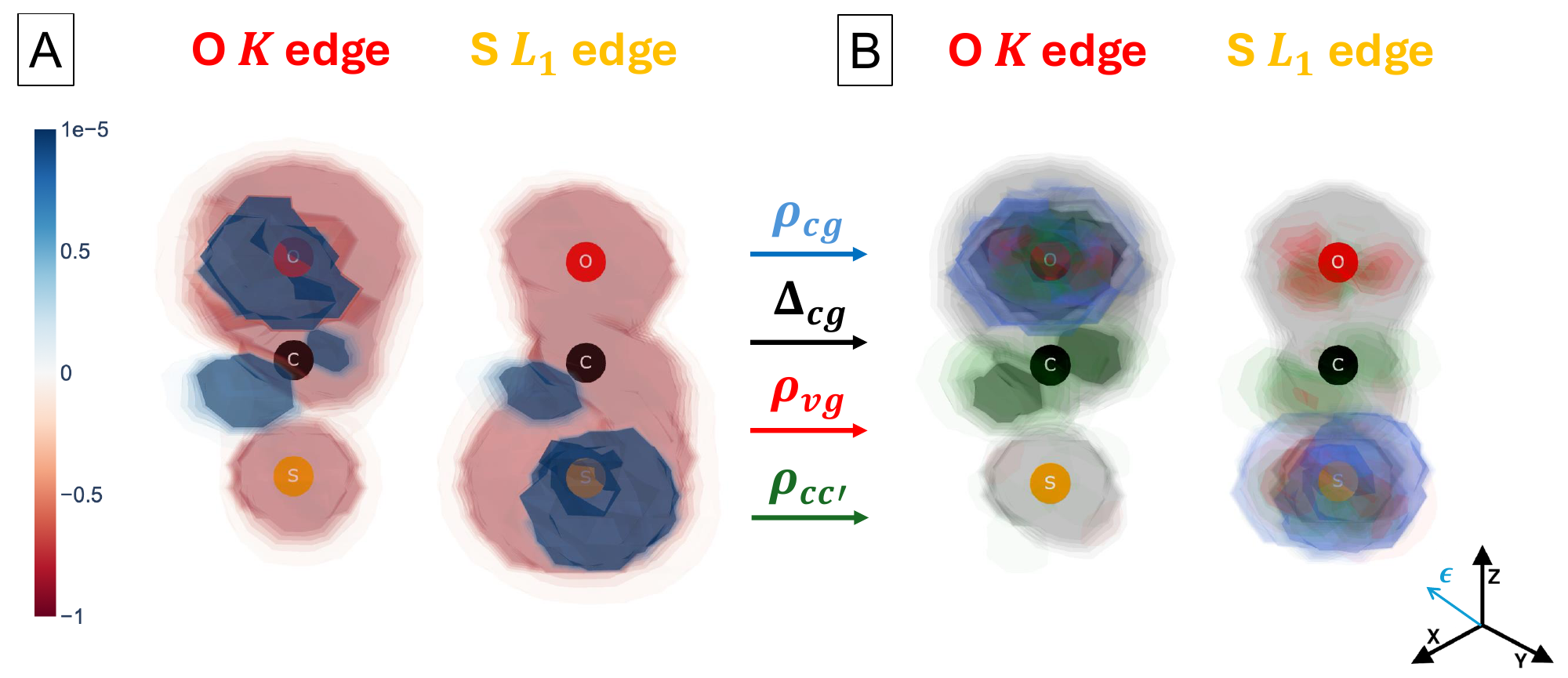}
    \caption{Difference density $\rho_{\Delta}(\bm{r}, t)$ and its decomposition into density components at 400 as after the peak of the x-ray pulse. (A) Snapshots of $\rho_{\Delta}(\bm{r}, t)$ at 400 as for excitation at the O $K$ edge and S $L_1$ edge with a pulse polarized at 45 degrees in the \textit{xz} plane. (B) Visual decomposition of $\rho_{\Delta}(\bm{r}, t)$ in terms of the contributions defined in \cref{eqn:density_comp}. The sketch in the lower right corner illustrates the orientation of the axes. The cyan vector shows the orientation of the x-ray polarization vector.}
    \label{fig:OCS_diff_dens_hierachy}
\end{figure*}

In \cref{fig:OCS_diff_dens_hierachy}A, we compare $\rho_{\Delta}(\bm{r},t)$ excited at the O $K$ edge (left) and at the S $L_1$ edge (right) with a pulse polarized at 45 degrees in the \textit{xz} plane. In particular, we compare snapshots at 400 as after the peak of the x-ray pulse. At this point in time, the pulse has already begun to fade and the \ac{wp} formed starts undergoing its dynamical evolution. In both cases, negatively-valued regions, corresponding to density reduction, are distributed over the whole molecule, whereas positively-valued regions, corresponding to density increase, are concentrated mainly at the pumped atom and the neighboring carbon. 

To explain this atom-specific distribution of $\rho_{\Delta}(\bm{r}, t)$, we decompose it into the contributions introduced in \cref{eqn:density_comp}, as shown in \ref{fig:OCS_diff_dens_hierachy}B. Each component contributes to different regions of $\rho_{\Delta}(\bm{r}, t)$, the contributions of $\rho_{cg}(\bm{r},t)$ and $\Delta_{cg}(\bm{r},t)$ generally being larger in magnitude than those coming from $\rho_{cc'}(\bm{r},t)$ and $\rho_{vg}(\bm{r},t)$ \footnote{This corresponds to quantitative differences in the relative magnitudes of the corresponding populations and coherences. In fact, with reference to \cref{eqn:density_comp}, $\rho^{(1)}_{cg}(t)$ is in the order $10^{-2}$, $\rho_{cc}^{(2)}(t)$ and $\rho_{gg}^{(2)}(t)$ are the order $10^{-3}$, while $\rho_{cc'}^{(2)}(t)$ and $\rho_{vg}^{(2)}(t)$ are in the order $10^{-4}$.}.

\begin{figure*}[t!]
    \centering
    \includegraphics[width=\textwidth]{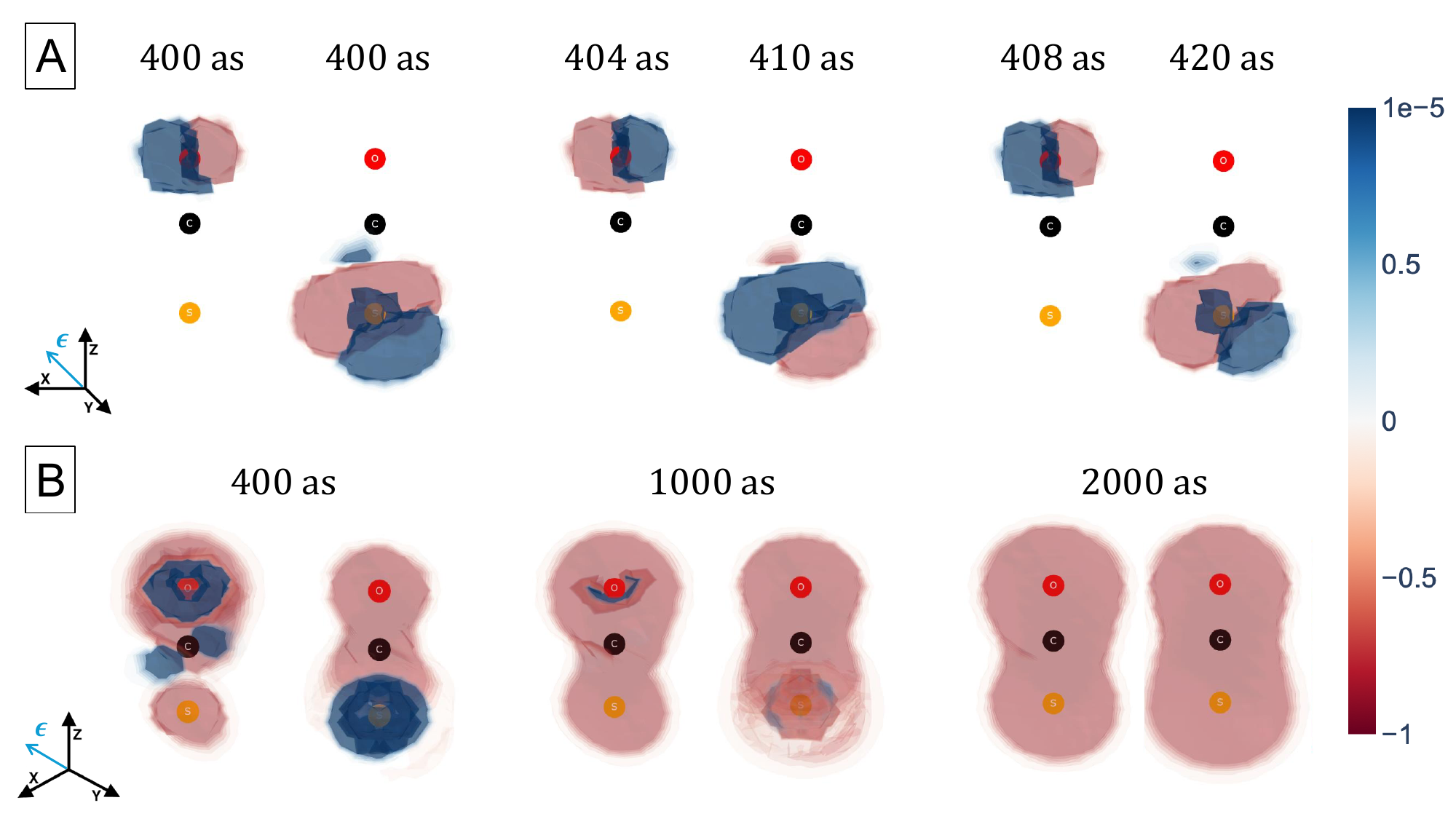}
    \caption{Time evolution of $\rho_{cg}(\bm{r},t)$ (A) and $\Delta_{cg}(\bm{r},t)$ (B) for a pulse polarized at 45 degrees in the \textit{xz} plane. In each panel, the snapshots are ordered in three groups, with time increasing from left to right. The left and right snapshots in each group correspond to the excitation at the O $K$ edge and S $L_1$ edge, respectively. The sketch in the lower left corner of each panel illustrates the orientation of the axes. The cyan vector corresponds to the orientation of the polarization vector.}
    \label{fig:OCS_time_evo_main_comp}
\end{figure*}

The $\rho_{cg}(\bm{r},t)$ component (blue) corresponds to the polarization in the electronic distribution established in linear response to the interaction with the x-ray pulse. We show the details of its time evolution in \cref{fig:OCS_time_evo_main_comp}A. At the O $K$ edge (left snapshots), $\rho_{cg}(\bm{r},t)$ sharply localizes on the oxygen atom and oscillates at a very high frequency proportional to the excitation energy at the O $K$ edge. At the S $L_1$ edge (right snapshots), $\rho_{cg}(\bm{r},t)$ localizes on sulfur and oscillates with a longer period than at the O $K$ edge, proportionally to the lower excitation energies at the S $L_1$ edge. The atom-specific localization of $\rho_{cg}(\bm{r},t)$ shows how the coupling between the x-ray field and the molecule takes place locally on the pumped atom; the linear response induced by the coupling remains localized on the pumped atom and oscillates at the frequency of the resonant transitions involved. 

Since $\bm{\epsilon}$ has components along both the $x$ and $z$ axes, excitations to both $\Pi_x$-symmetric and $\Sigma^+$-symmetric core-excited states are allowed. The orientation of $\rho_{cg}(\bm{r},t)$ in the \textit{xz} plane reflects the relative contributions from the two types of states. At the O $K$ edge, $\rho_{cg}(\bm{r},t)$ oscillates along the $x$ axis, signaling the dominant contribution of excitations to $\Pi_x$-symmetric states. At the S $L_1$ edge, $\rho_{cg}(\bm{r},t)$ is tilted in the \textit{xz} plane indicating significant contributions from both $\Pi_x$-symmetric and $\Sigma^+$-symmetric states.

The $\Delta_{cg}(\bm{r},t)$ component corresponds to the difference density obtained by subtracting the (time-independent) ground-state density from the (time-dependent) densities of the core-excited states. In the snapshots from the time evolution of $\Delta_{cg},(\bm{r},t)$ shown in \cref{fig:OCS_time_evo_main_comp}B, the negatively-valued regions extend over the whole molecule, completely offsetting the positively-valued regions as time proceeds. This behavior is caused by the limited lifetime of the core-excited states, which leads to the complete decay of their population within few femtoseconds. The positively-valued regions at 400 as, before significant decay has taken place, reflect the population of valence-type orbitals by the core-excited electron. At the O $K$ edge (left), the distribution resembles a $\pi^*$ orbital extending over the carbon and the oxygen; this shows how the
x-ray excitation can be described as a $1s\to\pi^*$ excitation. This is not the case at the S $L_1$ edge, where the x-ray excitation involves both $\Pi_x$-symmetric and $\Sigma^+$-symmetric states. As shown in the right snapshot at 400 as, the positively-valued region resides practically exclusively on sulfur, demonstrating a substantially local character of the excitation.

The $\rho_{cc'}(\bm{r},t)$ component corresponds to the coherences established between the core-excited states. Since the energy separation between core-excited states is only between 1 eV and 10 eV, the coherences associated to $\rho_{cc'}(\bm{r},t)$ oscillate with periods ranging between hundreds of attoseconds and few femtoseconds. The oscillation periods of the coherences translate directly to the time scale of the electron-density oscillations within $\rho_{cc'}(\bm{r},t)$. As shown in \cref{fig:OCS_diff_dens_hierachy}B, $\rho_{cc'}(\bm{r},t)$ provides an initial contribution to $\rho_{\Delta}(\bm{r},t)$, although to a lesser extent compared to $\rho_{cg}(\bm{r},t)$ and $\Delta_{cg}(\bm{r},t)$. Moreover, its effect on $\rho_{\Delta}(\bm{r},t)$ quickly fades away due to the fast decay of the core-excited states and is, therefore, not considered further in this work.

The $\rho_{vg}(\bm{r},t)$ component shown in \cref{fig:OCS_diff_dens_hierachy}B corresponds to the coherences established between the ground state and the valence-excited states via the \ac{isxrs} process. Its contribution is comparable to that of $\rho_{cc'}(\bm{r},t)$ in the initial stages of the excitation. However, owing to the much longer lifetime of the valence-excited states compared to the core-excited states \cite{Gamma_valence}, $\rho_{vg}(\bm{r},t)$ decays at a much slower rate than the other components. As time progresses and the core-excited components decay, $\rho_{vg}(\bm{r},t)$ gradually becomes the main contribution to $\rho_{\Delta}(\bm{r},t)$ (in addition to the stationary contribution from ground-state depopulation). Therefore, on time scales longer than the core-hole lifetime, charge migration within the molecule is captured by $\rho_{vg}(\bm{r},t)$. In the next subsection, we analyze in detail the electron dynamics within $\rho_{vg}(\bm{r},t)$, focusing, in particular, on the role of the x-ray polarization.

\newpage
\subsection{Charge migration in OCS}\label{sec:OCS_migration}
\begin{figure*}[h]
    \centering
    \includegraphics[width=\textwidth]{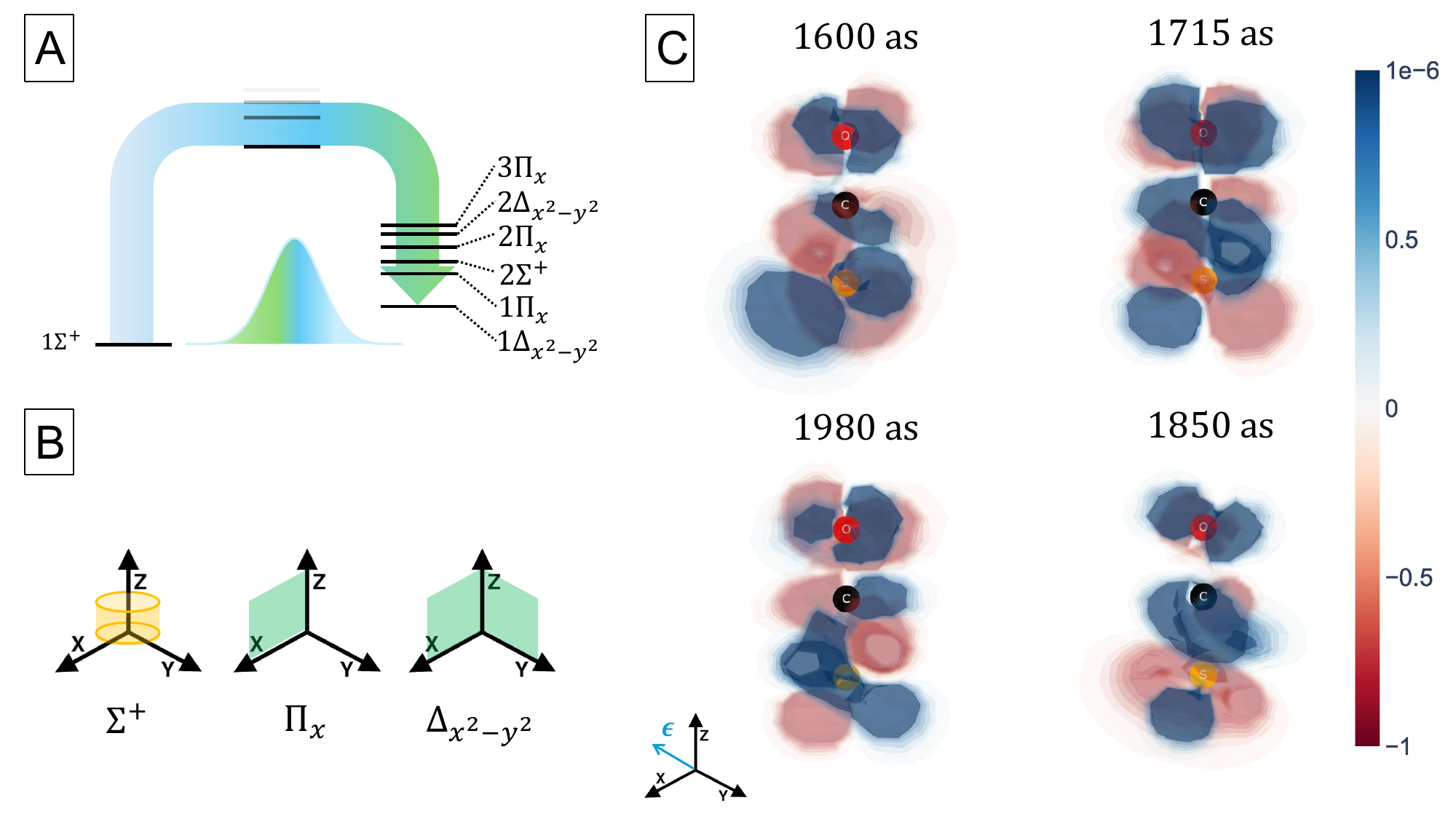}
    \caption{Properties of a valence-electronic wave packet excited by \ac{isxrs} at the S $L_1$ edge with an x-ray pulse polarized at 45 degrees in the \textit{xz} plane. (A) Illustration of \ac{isxrs} excitation of OCS with a broadband x-ray pulse. (B) Spatial symmetry of the $\Sigma^+$, $\Pi_x$, and $\Delta_{x^2-y^2}$ valence-excited states. The yellow cylinder represents the cylindrical symmetry of the $\Sigma^+$, while the green areas indicate the symmetry planes characterizing the $\Pi_x$ and $\Delta_{x^2-y^2}$ states. (C) Snapshots illustrating the charge migration by $\rho_{vg}(\bm{r}, t)$ over a 380 as period \cite{movies_repo}.}
    \label{fig:OCS_elec_migration}
\end{figure*}
The preparation of $\rho_{vg}(\bm{r},t)$ and the control over the associated charge migration relies on the tailored combination of molecular and pulse properties via the \ac{isxrs} process. In fact, based on the properties of the molecule---i.e., its energy spectrum and the dipole moments between its states---one can use the characteristics of the x-ray pulse (such as its bandwidth, central frequency, and polarization) to control the electronic \ac{wp}.

In the example presented in Fig. \ref{fig:OCS_elec_migration}, we considered a bandwidth of 8 eV (as already mentioned in \cref{sec:comp_details}) and a polarization oriented at 45 degrees in the \textit{xz} plane. As shown in \cref{fig:OCS_elec_migration}A, these pulse parameters allow us to substantially populate 5 valence-excited states of either $\Sigma^+$, $\Pi_x$, or $\Delta_{x^2-y^2}$ symmetry. The symmetry of the valence-excited states involved is reflected directly in the spatial distribution of the respective transition densities. In fact, as shown schematically in \cref{fig:OCS_elec_migration}B, the $\Sigma^+$ states are symmetric with respect to the \textit{z} axis, the $\Pi_x$ states are symmetric with respect to \textit{xz} plane, while the $\Delta_{x^2-y^2}$ states are symmetric with respect to both the \textit{xz} and the \textit{yz} planes. 

Since the electron dynamics within $\rho_{vg}(\bm{r},t)$ originate from the constructive and destructive interference of its coherent components, the spatial distribution of the transition densities directly influences the spatial properties of the charge-migration process, as illustrated in \cref{fig:OCS_elec_migration}C. A nodal plane bisecting the CO bond splits the distribution of $\rho_{vg}(\bm{r},t)$ in two parts, isolating the electron dynamics around the oxygen from those on the carbon-sulfur moiety. The electron dynamics involving carbon and sulfur are particularly pronounced. Starting from the snapshot at 1600 as, a lump of $\rho_{vg}(\bm{r},t)$ accumulates on sulfur on the left side of the \textit{yz} plane. The charge migration subsequently proceeds towards the right side of the \textit{yz} plane and then towards the carbon, as shown in the snapshot at 1715 as. As may be seen in the snapshot at 1850 as, the charge migration then changes its direction and proceeds around the carbon towards the left side of the \textit{yz} plane. Eventually, as shown in the snapshot at 1980 as, the electron density flows back to sulfur on the left side of the \textit{yz} plane. 

The charge migration thus follows a quasicircular path across the \textit{yz} plane, which breaks the symmetry with respect to the latter and preserves it with respect to the \textit{xz} plane. This depends directly on the symmetries of the states involved, with all states being symmetric with respect to the \textit{xz} plane and the $\Pi_x$ states providing the symmetry-breaking that enables the circular migration across the \textit{yz} plane. This example shows the connection between the symmetry of the states involved in the coherent superposition with the spatial pattern followed by the charge migration. Since the symmetry of states involved depends directly on the pulse polarization, tuning the latter makes it possible to exert a certain degree of control over the charge migration.

\newpage
\subsection{Charge migration in oxazole}\label{sec:oxazole_migration}
\begin{figure*}[h]
    \centering
    \includegraphics[width=\textwidth]{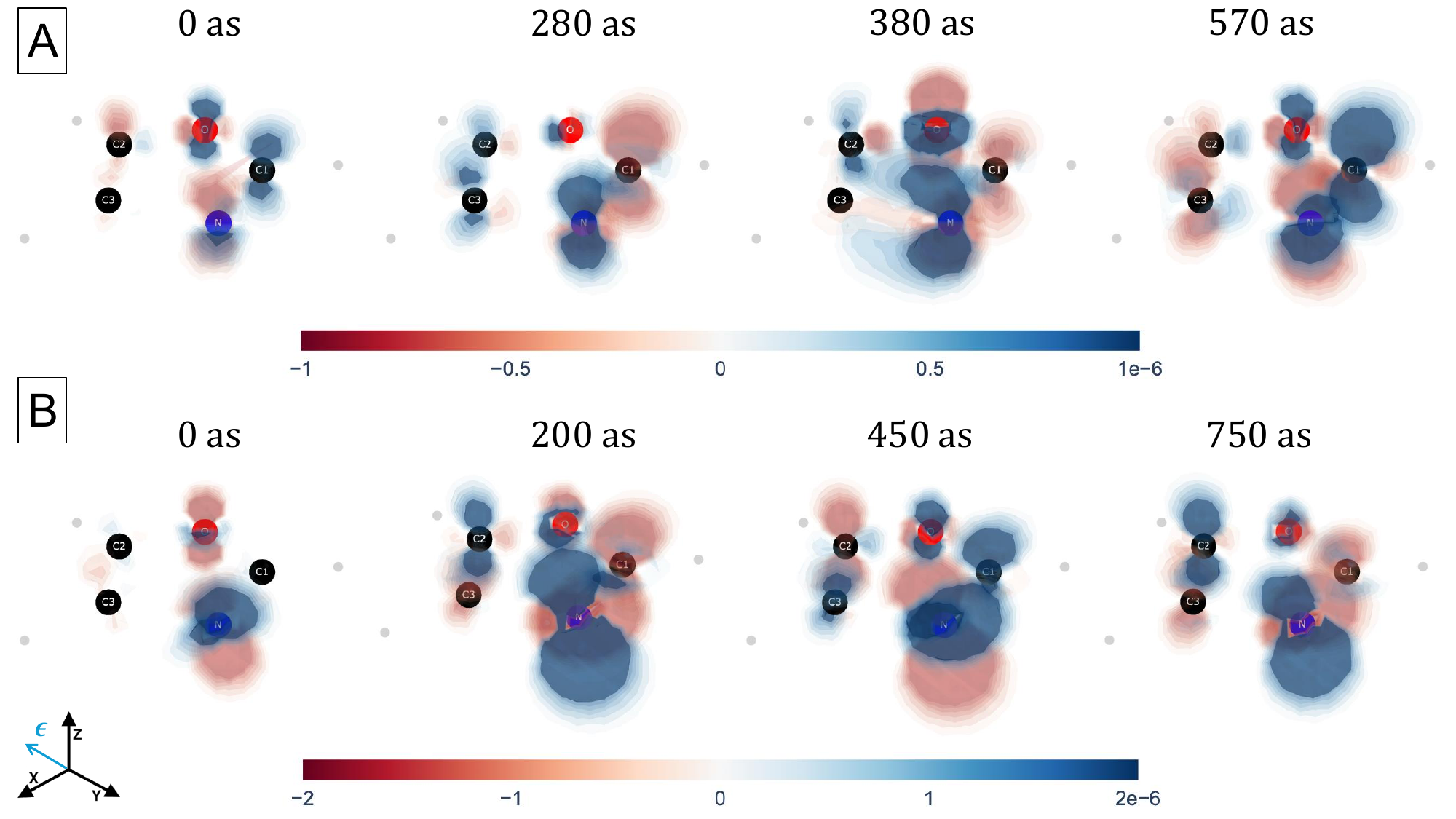}
    \caption{Atom-specific charge migration in oxazole, as captured by $\rho_{vg}(\bm{r},t)$, following the excitation with an x-ray pulse polarized at 45 degrees in the \textit{xz} plane. (A) Excitation at the O $K$ edge. (B) Excitation at the N $K$ edge. The sketch in the lower left corner shows the orientation of the axes. The cyan vector corresponds to the orientation of the polarization vector \cite{movies_repo}.}
    \label{fig:oxazole_elec_migration}
\end{figure*}
We applied the decomposition of $\rho_{\Delta}(\bm{r},t)$ into its perturbative components also to oxazole. This heterocyclic aromatic molecule contains two heteroatoms, oxygen and nitrogen, and in its equilibrium geometry presents a planar structure of $C_s$ symmetry. Here, we study the properties of $\rho_{\Delta}(\bm{r},t)$ excited either at the O $K$ edge or at the N $K$ edge. Similarly to OCS, in the initial stages of the time evolution $\rho_{cg}(\bm{r},t)$ and $\Delta_{cg}(\bm{r},t)$ provide the dominant contributions, their distribution displaying the atom-specific, local character of the x-ray excitation. The decay of the core-excited components gives way to $\rho_{vg}(\bm{r},t)$, which progressively becomes the main nonstationary contribution to $\rho_{\Delta}(\bm{r},t)$. 

The spectrum of oxazole includes several low-lying valence-excited states; their symmetry depends on the irreducible representation of the $C_s$ group they belong to: $\text{A}'$ states are symmetric upon reflection across the molecular plane (the $xy$ plane in Fig.~\ref{fig:oxazole_elec_migration}), while $\text{A}''$ states are antisymmetric. Using a pulse polarized at 45 degrees in the $xz$ plane both $\text{A}'$-symmetric and $\text{A}''$-symmetric states can be populated. In our case, using a pulse with an 8-eV bandwidth, we can substantially populate 20 valence-excited states (divided equally between the $\text{A}'$ and $\text{A}''$ symmetries) within the coherent superposition. Similarly to the example in OCS, including many states of different symmetries enables complex interference patterns between the corresponding coherent components, which, in turn, support rich electron dynamics. We present an example of the electronic dynamics corresponding to $\rho_{vg}(\bm{r},t)$ in Fig. \ref{fig:oxazole_elec_migration}, where we compare an excitation at the O $K$ edge (panel A) and at the N $K$ edge (panel B). 

The intensity build-up of the pulse corresponds to the preparation phase of $\rho_{vg}(\bm{r},t)$, during which its magnitude grows linearly and the electronic motion is frozen \cite{WP_brummer_Shapiro}. As the pulse gradually fades, the build-up of the coherence terminates and the electronic system evolves freely. As a first step, we analyze the electronic distribution at 0 as, which corresponds to the maximum pulse intensity and the peak of the \ac{wp} preparation. At this point in time, we can observe the atom-specific effects of the \ac{isxrs} excitation on the starting point of the charge migration. As shown in previous works \cite{Yong_Mukamel_2021,Fouda_2021,Balbi_Skeidsvoll_Koch_2023}, the atom-specific response to an \ac{isxrs} excitation is encoded in the relative amplitudes and phases of the coherent components of $\rho_{vg}(\bm{r},t)$. In our example, the atomic specificity is reflected directly in the initial distribution of $\rho_{vg}(\bm{r},t)$. In fact, when the molecule is excited at the O $K$ edge (panel A) the density accumulates predominantly on the oxygen and C1, whereas when the molecule is excited at the N $K$ edge (panel B), the density accumulates mainly on the nitrogen. This shows how the \ac{isxrs} process can be used to control the starting point of charge migration, which can be chosen by tuning the x-ray pulse to a specific inner-shell edge of the molecule.

Departing from the edge-dependent initial configuration, the charge migration initially unfolds in a clockwise direction. This can be seen by observing the density increase/decrease on C1, C2, and C3 in the snapshots in Fig. \ref{fig:oxazole_elec_migration}. At the O $K$ edge (panel A), the migration starts from O and C1 at 0 as, moves clockwise through N and C3 at 280 as, reaches C2 at 380 as, and returns to C1 at 570 as. At the N $K$ edge (panel B), starting from N at 0 as the charge migration reaches C2 at 200 as, quickly passes through C1 and C3 at 450 as, and eventually returns to C2 at 750 as. Following the first few hundred attoseconds \cite{movies_repo}, the subsequent electron dynamics behave similarly to clockwise and anticlockwise ring currents, as is typical for heteroaromatic systems of this type \cite{Elfl_Benzene_Tremblay_2016, QC_Benzene_Tremblay_Yang_2017, Mineo_Lin_Fujimura_2013, Moreno_Carrascosa_2021, Giri_Dixit_Tremblay_2023}. None of the directions prevails, with the two modes alternating---in a pattern characteristic of the chosen x-ray excitation edge---and the electron dynamics often being limited to oscillations between pairs of atoms.

Similarly to the OCS example, mixing states of $\text{A}'$ and $\text{A}''$ symmetry enables the charge migration to break through the symmetry plane of the molecule. This is not particularly evident at the O $K$ edge (panel A), where the $\text{A}'$-symmetric states are predominantly populated. However, at the N $K$ edge (panel B), owing to a comparable population of $\text{A}'$-symmetric and $\text{A}''$-symmetric states, the charge migration proceeds also across the $xy$ plane. This is particularly evident on the N atom, where the (positively- and negatively-valued) density lobes undergo a circular motion across the $xy$ plane.

\section{Conclusions}\label{sec:conclusions}
In this work, we presented a perturbative framework for computing the dynamics of an electronic \ac{wp} launched by an intense, broadband x-ray pulse in a neutral molecule. We coherently populated both the core-excited states and the valence-excited states by considering a linear x-ray absorption and a nonlinear \ac{isxrs} process as excitation mechanisms. We described the electronic structure and its properties at the \ac{eom-cc} level of theory, adopting an accurate implementation that avoids the truncation of the sum over states characterizing the \ac{rixs} transition moments. This provides an accurate description of the \ac{isxrs} process, improving on the commonly adopted truncation in the evaluation of the \ac{rixs} moments.

We studied the excitation and dynamics of an electronic \ac{wp} within OCS and oxazole, by monitoring the evolution of the time-dependent difference electron density. We separated the contributions of the excitation mechanisms by decomposing the time-dependent difference electron density into its perturbative components. We observe that the density components associated with core-excited states provide the main contributions to the difference density during the first few femtoseconds of the electronic dynamics. Their spatial distribution gives a visual representation of the atomic specificity and local character of the x-ray excitation. The relative contribution of the components changes with time: while the population of the core-excited states decays within few femtoseconds, the relative contribution of the valence-excited component progressively becomes the dominant nonstationary one. The coherent population of the valence-excited states via \ac{isxrs} provides a longer-lived platform for charge migration compared to the core-excited states. 

We focused more in depth on the valence-electron dynamics by studying the effect of the x-ray pulse parameters on the \ac{wp} launched via \ac{isxrs}. In both OCS and oxazole, we showed how the symmetry of the states included in the quantum superposition shapes the spatial pattern followed by charge migration. Owing to its direct relation with the symmetry of the states (mediated by electric-dipole selection rules), the pulse polarization can provide a degree of control over charge migration and its spatial properties. In oxazole, we showed how the atom specificity of ISXRS is reflected in the initial distribution of the electron density. This shows the capability of \ac{isxrs} to spatially control the starting point of charge migration by tuning the x-ray pulse to a specific inner-shell edge of the molecule. We examined the characteristics of the subsequent charge migration, showing how it unfolds according to a well-known pattern of clockwise and anticlockwise ring currents. 

In this paper, we studied purely coherent electron dynamics without considering the decoherence effects associated with the nuclear degrees of freedom \cite{nuc_dec_Vacher_2015,nuc_dec_Vacher_2017,nuc_dec_Arnold_2017}. Given the absence of hydrogen atoms in OCS, we don't expect that the inclusion of ground-state structural fluctuations would lead to considerable modifications of the electron dynamics in OCS that we have predicted. In oxazole, structural quantum fluctuations are more likely to have an impact on electronic coherence, but not so much on the sub-fs time scale we considered in Fig.~\ref{fig:oxazole_elec_migration}. Nevertheless, an important future extension of this work entails including such effects in the theoretical framework. This would provide a more complete picture of the excitation, the intramolecular electron dynamics, and their dependence on the x-ray pulse parameters, while extending the capabilities of the framework towards fully predictive calculations of attosecond-pulse-induced chemical dynamics.

\clearpage
\begin{acknowledgments}
We thank Dr. Mads Pedersen for his help with the computational implementation. This work was supported by the Cluster of Excellence
'CUI: Advanced Imaging of Matter' of the Deutsche Forschungsgemeinschaft (DFG)
– EXC 2056 – project ID 390715994. E. R., S. C., N. R., and R. S. acknowledge support from DESY (Hamburg, Germany), a member of the Helmholtz Association HGF. The work at USC has been supported by the US National Science Foundation (grant CHE-2525964). A.I.K. also acknowledges support from the Mildred Dresselhaus Guest Professorship during her sabbatical stay at DESY.
\end{acknowledgments}

\appendix
\section{Optimized nuclear geometries}\label{App:equilibrium_geom}
\begin{table}[h]
    \caption{Equilibrium geometry of OCS.}
    \label{tbl:eq_geometries_ocs}
    \begin{tabular}{l *{3}{S[table-format=-1.3]}}
        \toprule
        Atom & {$x$ (\AA)} & {$y$ (\AA)} & {$z$ (\AA)} \\
        \midrule
        S    & 0.000       & 0.000       & -1.052      \\
        C    & 0.000       & 0.000       &  0.537      \\
        O    & 0.000       & 0.000       &  1.701      \\
        \bottomrule
    \end{tabular}
\end{table}

\begin{table}[h]
    \caption{Equilibrium geometry of oxazole.}
    \label{tbl:eq_geometries_oxazole}
    \begin{tabular}{l *{3}{S[table-format=-1.3]}}
        \toprule
        Atom & {$x$ (\AA)} & {$y$ (\AA)} & {$z$ (\AA)} \\
        \midrule
        O    & -0.360      & -1.096      &  0.000      \\
        C    & -1.105      &  0.047      &  0.000      \\
        H    & -2.185      & -0.074      &  0.000      \\
        N    & -0.422      &  1.154      &  0.000      \\
        C    &  0.944      & -0.644      &  0.000      \\
        H    &  1.726      & -1.397      &  0.000      \\
        C    &  0.918      &  0.720      &  0.000      \\
        H    &  1.750      &  1.420      &  0.000      \\
        \bottomrule
    \end{tabular}
\end{table}

\section{Quantum-chemical calculation protocol}\label{App:calc_protocol}
The electronic-structure calculation follows a 3-step protocol: a restricted \ac{hf} calculation precedes a ground-state CCSD calculation, with the final step being an fc-CVS-EOM-EE-CCSD calculation. In order to assemble the time-dependent electron density correctly, it is necessary to obtain all the necessary quantities from a single calculation. This is needed since each \ac{hf} molecular orbital and \acs{eom-cc} state is determined only up to a sign, which may, in principle, change between different instances of the same calculation. Assembling the electron density using the output of a single calculation instance ensures a consistent treatment of the signs and the reproducibility of the results. 

\section{Energy spectra}\label{App:OCS_spectrum}
\subsection{Energy spectrum of OCS}
The transition energies relative to the ground state $1\Sigma^+$ for the considered valence-excited states are shown in Table \ref{table:OCS-ve_data}. The states are labeled according to the irreducible representations of the $C_{\infty v}$ group. The transition energies relative to the ground state $1\Sigma^+$ for the considered core-excited states are shown in Table \ref{table:OCS-ce_data}. The $\omega_{cg}^S$ columns correspond to the excitation energies at the S $L1$ edge, the $\omega_{cg}^O$ columns correspond to the excitation energies at the O $K$ edge. The values highlighted in bold correspond to the calculated energies for the $1s\to\pi^*$ transition.

\begin{table}[h]
    \caption{Spectrum of valence-excited states of OCS.}
    \label{table:OCS-ve_data}
    \begin{tabular}{l S[table-format=1.2]} 
        \toprule
        $\ket{\psi_v}$ & {$\omega_{vg}$ (eV)} \\ 
        \midrule
        $1\Delta_{x^2-y^2}$ & 5.75 \\
        $1\Pi_x$            & 7.21 \\
        $2\Sigma^+$         & 8.22 \\
        $2\Pi_x$            & 8.43 \\
        $2\Delta_{x^2-y^2}$ & 9.10 \\
        $3\Pi_x$            & 9.18 \\
        \bottomrule
    \end{tabular}
\end{table}

\begin{table}[h]
    \caption{Spectrum of core-excited states of OCS.}
    \label{table:OCS-ce_data}
    \begin{tabular}{l S[table-format=3.2] S[table-format=3.2] @{\hspace{2em}} l S[table-format=3.2] S[table-format=3.2]}
        \toprule
        $\ket{\psi_c}$ & {$\omega_{cg}^S$ (eV)} & {$\omega_{cg}^O$ (eV)} & $\ket{\psi_c}$ & {$\omega_{cg}^S$ (eV)} & {$\omega_{cg}^O$ (eV)} \\ 
        \cmidrule(r{2em}){1-3} \cmidrule{4-6}
        $1^c\Sigma^+$  & 230.74                 & 538.50                 & $1^c\Pi_x$, $1^c\Pi_y$ & {\textbf{229.74}} & {\textbf{535.63}} \\
        $2^c\Sigma^+$  & 232.07                 & 540.61                 & $2^c\Pi_x$, $2^c\Pi_y$ & 233.50            & 540.84            \\
        $3^c\Sigma^+$  & 232.42                 & 541.32                 & $3^c\Pi_x$, $3^c\Pi_y$ & 234.98            & 542.42            \\
        $4^c\Sigma^+$  & 235.00                 & 541.89                 & $4^c\Pi_x$, $4^c\Pi_y$ & 239.75            & 545.63            \\
        $5^c\Sigma^+$  & 236.54                 & 543.97                 & $5^c\Pi_x$, $5^c\Pi_y$ & 240.48            & 549.20            \\
        \bottomrule
    \end{tabular}
\end{table}

\subsection{Energy spectrum of oxazole}\label{SIsec:Oxazole_spectrum}
The transition energies relative to the ground state $1\text{A}'$ for the considered valence-excited states are shown in Table \ref{table:Oxazole-ve_data}. The states are labeled according to the irreducible representations of the $C_s$ group. The transition energies relative to the ground state $1\text{A}'$ for the considered core-excited states are shown in Table \ref{table:OXAZOLE-ce_data}. The $\omega_{cg}^N$ columns correspond to the excitation energies at the N $K$ edge, the $\omega_{cg}^O$ columns correspond to the excitation energies at the O $K$ Edge. The values highlighted in bold correspond to the calculated energies for the $1s\to\pi^*$ transition.
\begin{table}[h]
    \caption{Spectrum of valence-excited states of oxazole.}
    \label{table:Oxazole-ve_data}
    \begin{tabular}{l S[table-format=1.2] @{\hspace{2em}} l S[table-format=1.2]} 
        \toprule
        $\ket{\psi_v}$ & {$\omega_{vg}$ (eV)} & $\ket{\psi_v}$ & {$\omega_{vg}$ (eV)} \\ 
        \cmidrule(r{2em}){1-2} \cmidrule{3-4}
        $2\text{A}'$  & 6.76 & $1\text{A}''$  & 6.78 \\
        $3\text{A}'$  & 7.42 & $2\text{A}''$  & 6.97 \\
        $4\text{A}'$  & 8.18 & $3\text{A}''$  & 7.57 \\
        $5\text{A}'$  & 8.31 & $4\text{A}''$  & 7.90 \\
        $6\text{A}'$  & 8.80 & $5\text{A}''$  & 8.06 \\
        $7\text{A}'$  & 8.99 & $6\text{A}''$  & 8.41 \\
        $8\text{A}'$  & 9.23 & $7\text{A}''$  & 8.57 \\
        $9\text{A}'$  & 9.33 & $8\text{A}''$  & 8.76 \\
        $10\text{A}'$ & 9.48 & $9\text{A}''$  & 8.82 \\
        $11\text{A}'$ & 9.68 & $10\text{A}''$ & 9.45 \\
        \bottomrule
    \end{tabular}
\end{table}

\begin{table}[h]
    \caption{Spectrum of core-excited states of oxazole.}
    \label{table:OXAZOLE-ce_data}
    \begin{tabular}{l S[table-format=3.2] S[table-format=3.2] @{\hspace{2em}} l S[table-format=3.2] S[table-format=3.2]}
        \toprule
        $\ket{\psi_c}$ & {$\omega^{N}_{cg}$ (eV)} & {$\omega^{O}_{cg}$ (eV)} & $\ket{\psi_c}$ & {$\omega^{N}_{cg}$ (eV)} & {$\omega^{O}_{cg}$ (eV)} \\ 
        \cmidrule(r{2em}){1-3} \cmidrule{4-6}
        $1^c\text{A}'$  & 403.99 & 539.16 & $1^c\text{A}''$  & {\textbf{400.87}} & {\textbf{536.99}} \\
        $2^c\text{A}'$  & 404.73 & 540.10 & $2^c\text{A}''$  & 403.83            & 540.32            \\
        $3^c\text{A}'$  & 404.97 & 540.29 & $3^c\text{A}''$  & 405.13            & 540.79            \\
        $4^c\text{A}'$  & 405.01 & 540.67 & $4^c\text{A}''$  & 406.34            & 541.98            \\
        $5^c\text{A}'$  & 405.49 & 540.99 & $5^c\text{A}''$  & 406.97            & 542.66            \\
        $6^c\text{A}'$  & 405.88 & 541.34 & $6^c\text{A}''$  & 408.90            & 545.07            \\
        $7^c\text{A}'$  & 407.08 & 542.17 & $7^c\text{A}''$  & 410.95            & 545.74            \\
        $8^c\text{A}'$  & 407.29 & 542.73 & $8^c\text{A}''$  & 417.32            & 554.12            \\
        $9^c\text{A}'$  & 407.71 & 542.82 & $9^c\text{A}''$  & 418.02            & 554.36            \\
        $10^c\text{A}'$ & 408.33 & 543.46 & $10^c\text{A}''$ & 418.93            & 556.28            \\
        $11^c\text{A}'$ & 408.51 & 543.96 & $11^c\text{A}''$ & 419.58            & 556.54            \\
        $12^c\text{A}'$ & 408.65 & 544.58 & $12^c\text{A}''$ & 420.49            & 559.63            \\
        $13^c\text{A}'$ & 409.00 & 544.76 & $13^c\text{A}''$ & 422.00            & 560.78            \\
        $14^c\text{A}'$ & 409.57 & 545.04 & $14^c\text{A}''$ & 422.39            & 560.99            \\
        $15^c\text{A}'$ & 409.99 & 545.88 & $15^c\text{A}''$ & 423.52            & 561.15            \\
        \bottomrule
    \end{tabular}
\end{table}

\bibliography{references}

\end{document}